# Emitting solitonized laser beams to boost the negative energy density of squeezed regions of the vacuum

*Mohammad Mansouryar*

**Abstract**: There are two main approaches to generate the negative energy density (NED) in the literature. The Casimir effect and the squeezed vacuum. The possibility of the latter approach is qualitatively reviewed in this paper. It is proposed that the soliton theory may give remarkable contributions to generate and separate the NED out of the quantum vacuum. By applying the wavelength-division multiplexing (WDM) method on the solitonized laser beams which give rise the squeezed regions of the vacuum, it is shown the solitons' properties can be useful to the chosen approach.

## I. Introduction

The only problem of generating those spacetime structures where effectively allow faster-than-light motions, is the needed exotic matter [1]. There are two main methods within the realm of quantum physics containing the negative energy (NE) in some states: The Casimir effect [2] and the squeezed vacuum [3, pp 476-479]. The prime energy requirement of the desired mentioned spacetimes is to have the NED in the free space in which one could engineer the associated metric [4]. The Casimir effect is able to give the NE inside the appropriate cavities [2]. However there has been suggestions to stretch those NE regions out of the walls [5, 6]; one observes the NED of Casimir method is intrinsically available in the *confined* configurations of the cavities [7]. However the NED related to the squeezed vacuum states is intrinsically available in the *free* space [3, pp 476-479; 8]. It is straightforward if one tries to separate the needed NED of squeezed vacuum and conduct it to the arbitrary location of the spacetime to reach to the geometry of a spacewarp.

There are three proposals, graphically described in [3, 9] for such style of separation, one of them implies "a set of rapidly rotating mirrors in which the light beam would be set to strike each mirror surface at a very shallow angle while the rotation would ensure that the negative energy pulses would be reflected at a slightly different angle from the positive energy pulses [3, 9]":

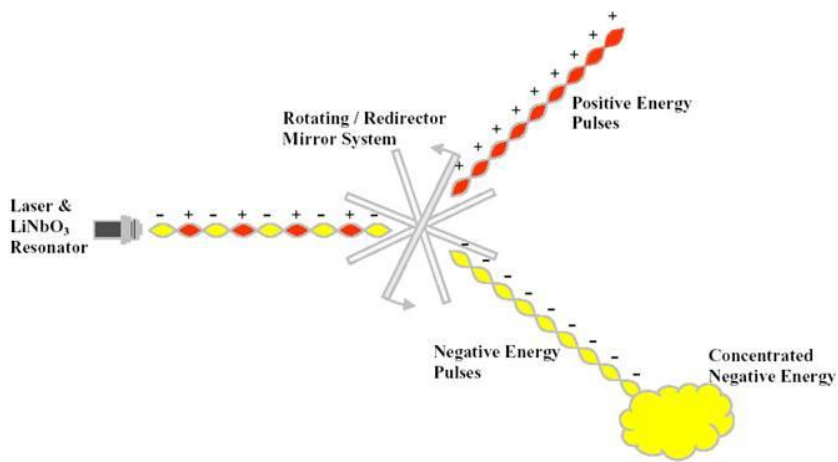

Figure 1

Altogether, the details of the setups in page 478 of [3] on exact mechanism of separation is ambiguous. For example, since in that concept both the negative and positive energy (PE) pulses are $\sim 10^{-15}$ second in duration, it is not known how an array of mirrors could cause separation between those pulses [10].

Anyway, if the objective is to separate the NE parts of the squeezed light, the idea could be conducting them to a course in which their summation could fill out the targeted spot of the spacetime to result the spacewarp metric [5]. This paper suggests the approach of using the solitons [11] to that end.

Since the soliton pulses can be generated by the laser arrangements [12], regarding the properties of a soliton wave, if one could cast a laser-made squeezed-state-shape into a soliton solution, possibly it is assumed to preserve the NE part of the beam for a longer distance and time either.

## II. The Squeezed Vacuum

Since the quantum noise is the uncertainty of some physical quantity due to its quantum origin, if one could decrease the noise in one observable, for instance energy, the Heisenberg uncertainty principles implies increasing the noise in the conjugate observable. Having applied this on the quantum vacuum, one may reduce the quantum fluctuations to create the NE states, viz "to squeeze" the vacuum.

Such squeezed states of the electromagnetic field can be generated by nonlinear optical processes, such as four-wave mixing, in which photons are created in pairs [13], a degenerate three-level laser containing a parametric amplifier and coupled to a squeezed vacuum reservoir [14], or a degenerate three-level cascade laser coupled to a vacuum reservoir via one of the coupler mirrors and an external resonant coherent radiation in the other [15]. It can be shown that in the model of a linear cavity Raman amplifier, and in arbitrary values of the experimentally related parameters, such as pump amplitude, line widths and dispersion, at any given time $t$, it is possible to choose optimal mode coefficients in such a way, that one of the quadratures has minimal variance below the vacuum level at that time; i.e., for which maximal squeezing and minimal state-reduction [13]. Indeed, by illustrating the sequence of squeezing and "anti-squeezing" operations performed by the cavity Raman laser (CRL) and the non-degenerate linear parametric amplifier (LPA), the authors of [13] state: "Experimentally, all of these parameters can be controlled independently, by varying the pressure of the gaseous Raman medium, the type of medium and the pump intensity"; and "There exist optimal linear combination modes, for which the variance of one quadrature is always below the vacuum value. Note the conventional definition of squeezing is that the variance of one of the Hermitian components is below the vacuum value $\frac{1}{2}$. The degree of decoherence for the combination mode is much less than for the individual Stokes and anti-Stokes modes".

For example, as a squeezing operation performed on the modes $b_1$ and $b_2$, implemented by the non-degenerate LPA:

$$c_1^{\dagger} = U^* b_1^{\dagger} - V^* b_2 \qquad c_2 = -V b_1^{\dagger} + U b_2 \qquad (1)$$

, where $b_1$ and $b_2$ are the linear combinations modes of uncorrelated normal modes $c_1$ and $c_2$; $|U|^2 - |V|^2 = 1$, see [13, sections 2-1 and 2-2].

One solution for propagation in the $z$-direction and the zero photon count for mode is given by:

$$c_2(z) = e^{i\eta} \sinh gz\, b_1^\dagger + \cosh gz\, b_2 \qquad (2)$$

By analyzing their work, it can be deduced that various innovations about squeezing are possible, if one manipulates the involved coefficients elegantly. As an example, for the case of hypertransient limit which the eigenvalues are all purely imaginary and the linear CRL system shows quasi-periodic behavior, minimal variance is found for conventional two-mode squeezing:

$$\varphi_{conv}^2 = \varphi_{min}^2 = \frac{1}{2}\left(\frac{\kappa_2 - \kappa_1}{\kappa_2 + \kappa_1}\right)^2 \qquad (3)$$

, which $\kappa_1$ and $\kappa_2$ are the coupling coefficients for the medium with the Stokes and anti-Stokes variables. In both cases, $\varphi_{min}^2$ and $\varphi_{fix}^2$ go to different values (less than $\frac{1}{2}$) as $t \to \infty$.

As another example, for the steady-state case that hopefully keeps us away from periodic behaviors, we have:

$$\varphi_{conv}^2 \to \frac{1}{2}\left(\frac{\kappa_2}{\kappa_2 + \kappa_1}\right)^2 < \frac{1}{2} \qquad (4)$$

Obviously, a more denominator, a more squeezed vacuum. These relations express that to optimize the amount of noise cancellation, the mode coefficients have to be chosen appropriately [16].

Almost perfect squeezing can be achieved at the interested option of steady state and at threshold for a suitable choice of parameters; in one case the intracavity squeezing is 97.8% below the vacuum level. Although, the models of [14] depend on several factors; e.g., the degree of squeezing of the of the output mode in the presence of the nonlinear crystal is greater by 1% than that without the nonlinear crystal, along with the presence of the squeezed vacuum reservoir increases the degree of squeezing significantly over and above the squeezing attainable from a degenerate three-level laser with a parametric amplifier [17], and for $\beta \neq 0$, where $\beta = \Omega/\gamma$, and $\Omega$ is proportional to the amplitude of the pump mode, and $\gamma$ is the atomic decay rate assumed to be the same for all the three levels [14]. Besides, the plots of Ref. [14] show that the degree of squeezing of the output mode increases with the squeeze parameter $r$ and almost perfect squeezing occurs for small values $\beta$.

It should be noted that the mean photon number of the cavity and output modes obtained at [14] or in a case that the ASDM soliton corresponds to the second function belonging to the set that extracts their stability features at three path-average dispersion circumstances and the third-order dispersion gives rise to an asymmetry of the DM-soliton's profile and generation of radiation [18], can be taken into account for further applications of the planned spacetime; specifically the PE contributions – in the form of squeezed regions and photons – can be applied to play the role of a "ballast" factor [19] for the Arnowitt-Deser-Misner (ADM) mass of the obtained spacetime [4; pp 111-113] to yield a controllable loop of losing mass of the emitting mouth in order to have much amount of NED from a quantum-scale effect. Also, the PE contributions could probably be deflected by metamaterial structures [20].

Another situation where it is believed that the system under consideration can generate an intense squeezed light is a degenerate three-level cascade laser, studied in [15]. Although the analysis is subjected to mathematical simplifications, it still gives interesting results for our purposes. In particular, a maximum of 98% squeezing (up to 98.3%, pertaining to the initial preparation of the superposition and strength of the coherent radiation) occurs at $\eta = 0.02$ for $A = 1000$; where $-1 \leq \eta \leq 1$ defined in Eqs. (42-46) of [15] and $A = 2r_a g^2 / \gamma^2$ ($r_a$ denotes the constant rate of injecting the three-level atoms initially prepared in a coherent superposition of the top and bottom levels into a cavity, $g$ is the coupling constant, taken to be the same for both transitions, and $\gamma$ represents the atomic spontaneous decay rate, taken to be the same for the two upper levels, for convenience). The correlated emission initiated by the initial atomic coherence is responsible for the reduction of the fluctuations of the noise in one of the quadrature components below the classical limit [15]. Additionally, the degree of squeezing would be maximum for certain values of the initial atomic coherence that depends on the rate at which the atoms are injected into the cavity; and for our interesting steady state case, the atomic variables will reach it in relatively short time. As Tesfa declares, it is also possible to notice that the more there are atoms initially in the upper level (are initially prepared with a maximum atomic coherence and also driven externally with a coherent radiation of relatively small amplitude), the better would be the resulting squeezing, – in the absence of the driving radiation, an intense radiation with high squeezing can be produced specially when the atoms are initially prepared with equal probability of being in the bottom and top levels where there is no squeezing, so driving mechanism can be considered as an option for producing a squeezed light when it is difficult to prepare the atoms in an arbitrary initial superposition [21] – ; though it is observed that pumping the atoms with stronger radiation than required considerably reduces the squeezing and intensity of the cavity radiation [15]. However, since the intensity of the produced light decreases with the strength of the external coherent light if the atoms are initially prepared to have equal probability of being in the top and bottom levels for $\Omega < \gamma$, so we should not manipulate the system externally in a gross manner and it would be more beneficial to let the interactions continue internally. Hence, to show the constraints, while at the Fig. 3 of [15], the depicted curve rapidly passes from squeezed state to upper states, we cannot use arbitrary values of $A$, since the steady state consideration fails to be applied for $A > 0.99$, for $\eta = 1$ and $\Omega = 3.5\gamma$.

If figures analysis continues, from the Fig. 5 of [15], one can observe an initial diving of the quadrature variance $\Delta a^2$ of the cavity radiation at steady state, toward squeezing values, while the curve brings itself out rather fast. Therefore, the required feat is providing a situation in which the system mostly stays in the diving-to-squeezing state, by engineering the involved parameters¶. One might use homodyne detection method to watch the story.

## III. The Solitonization Mechanism

After pondering on the figures about the squeezed states of the vacuum, here it comes the core idea: what if the below Fig. 4.a from [13] recited below, reminds one an optical soliton with a hyperbolic-secant envelope shape?

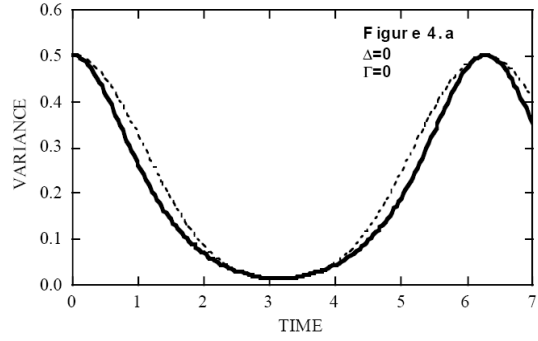

Figure 2

Obviously, the shape of the variance-time diagram is similar to a soliton. By changing the involved coefficients, it is interesting if one can reformulate it as a pulse envelope of a soliton function given as:

$$u(x,t) = -3c \, \text{sech}^2 \left[ c^{\frac{1}{2}}(x - ct)/2 \right] \quad (5)$$

The above function is a soliton solution of the Korteweg-deVries (KdV) equation [11]:

$$u_t = u_{xxx} + uu_x \quad (6)$$

, with an additional negative sign; for the N-soliton solution refer to [22, 23]. It is noteworthy to investigate if one could consider it as an antikink too. As it will be brought up later, it is interested to follow the soliton theory patterns and form the Fig. 2 into a soliton shape by adjusting the related variables [24], to be able to send a solitonized pulse containing the NE [25]. For such a system, we can use the idea of the WDM. The idea of WDM is that solitons are encoded on different wavelengths which will propagate with different velocities due to the dispersion of the fiber [26], and laser beam medium in present application (see section **IV**).

To cite several properties of the solitons regarding [23], notice the KdV solitary waves could be shown numerically to remain shape invariant upon interaction (undergoing only a phase shift), and interact elastically, and if more than one soliton is present in a given solution, then these solitons must be of different heights and travel at different constant speeds; so due to Special Relativity none of these solitons allow to move faster than the speed of light, if one could consider them as an information carrier. The effects of our relativistic situation (laser solitons) on the involved phase velocities, group velocities, front velocities, and signal velocities may be a matter of further research. Also when solitons collide and overlap, the linear superposition principle does not apply, so the KdV is scale invariant, which means we can set the value of the coefficients of its three terms to arbitrary values by choosing appropriate scaling transforms for $x$, $t$ and $u$; and only one unique solution is possible from an arbitrary initial soliton condition; besides, the KdV admits solitary waves travelling to the left or to the right (but not in both directions at once) for a given KdV evolution equation, in a way to avoid the oscillatory solution. The form of the one soliton solution is symmetrical around the point of maximum amplitude and must be an even function in $\eta$ which $\eta = px - \Omega t$ is a moving Galilean reference frame where $v = \frac{\Omega}{p}$ and $p$ is a free parameter.

For instance, considering the two-soliton solution by Eq. (2.86) of [23], describing two solitons travelling to the right, and the speed of each soliton is related to its amplitude; so we see that solitons must be of different height for a non-trivial solution[§]. Hence, in the two-soliton solution the taller soliton will travel faster than the smaller one, eventually overtaking it.

Having these interesting properties on the table, it may be significant to think about applying them on the quantum interest conjecture [27]. Can one manipulate a solitonized laser pulse in which the tall soliton would be the squeezed region of vacuum and the short soliton be the PE density (PED) one? What if the small regions of PE keep being soliton until the end and not to interact linearly in order to prevent entering a big amplified PE into the arrangement, while by altering the parameters, the large NE regions lose being in a soliton state after passing a duration, to behave classically for interacting and amplifying (enhancing themselves). This event can be realized by intervention a negative pulse in middle of the way[†] [28].

Another significant situation for this paper's applications is the Fig. 4c of [13] at left below, compared to Fig. 3 of [31]:

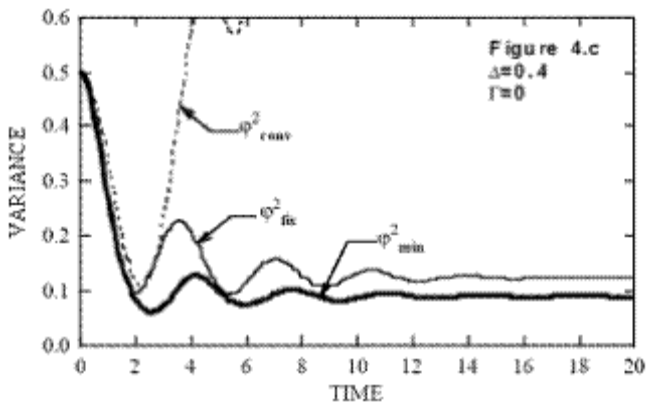 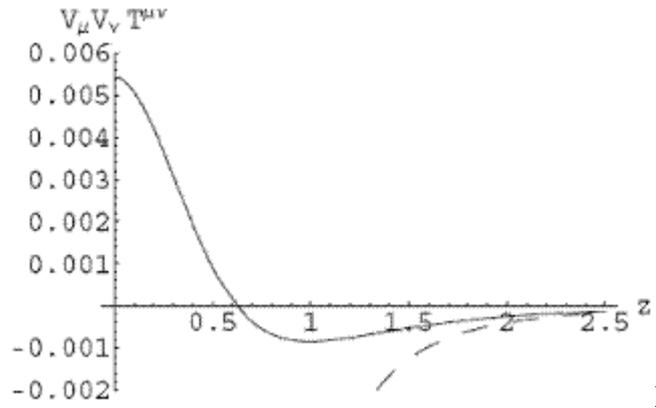

Figure 3

As an intriguing occurrence, one can find similarities between the behavior of the left above figure and the right one, that is the contributions to null energy condition in three dimensions for a Dirichlet plate with a hole of unit radius, as functions of distance along the axis passing through the center of the hole of a symmetrically perforated Casimir cavity [31]. *Remark:* The dotted lines of the right figure show the perfect mirror result.

The analogy is between the diagram of the variance of the minimal quadrature variable for the choice of minimizing mode coefficient $\varphi_{\min}^2$ (see Eq. (3) too) at zero linewidth ($\Gamma = 0$), for amount of dispersive detuning as $\Delta = 0.4$ [13], and the energy-momentum tensor for null vectors of a static and symmetric system as $T_{\lambda\nu}V^\lambda V^\nu = (V^\alpha \partial_\alpha \phi)^2 = \dot{\phi}^2 + \sum_i (v_i \partial_i \phi)^2 = \dot{\phi}^2 + (\partial_z \phi)^2$ ... $\implies \langle \dot{\phi}^2 \rangle + \langle (\partial_z \phi)^2 \rangle = -\frac{1}{16\pi^2 z^4}$ in three dimensions with Dirichlet conditions [31].

In this spirit, the mentioned feat of "maintaining the system in the dive-to-squeezing state, by engineering the involved parameters[¶]", can be imposed with the possible arrangements of the solitonization mechanism to yield the desired pulses, as graphically outlined below:

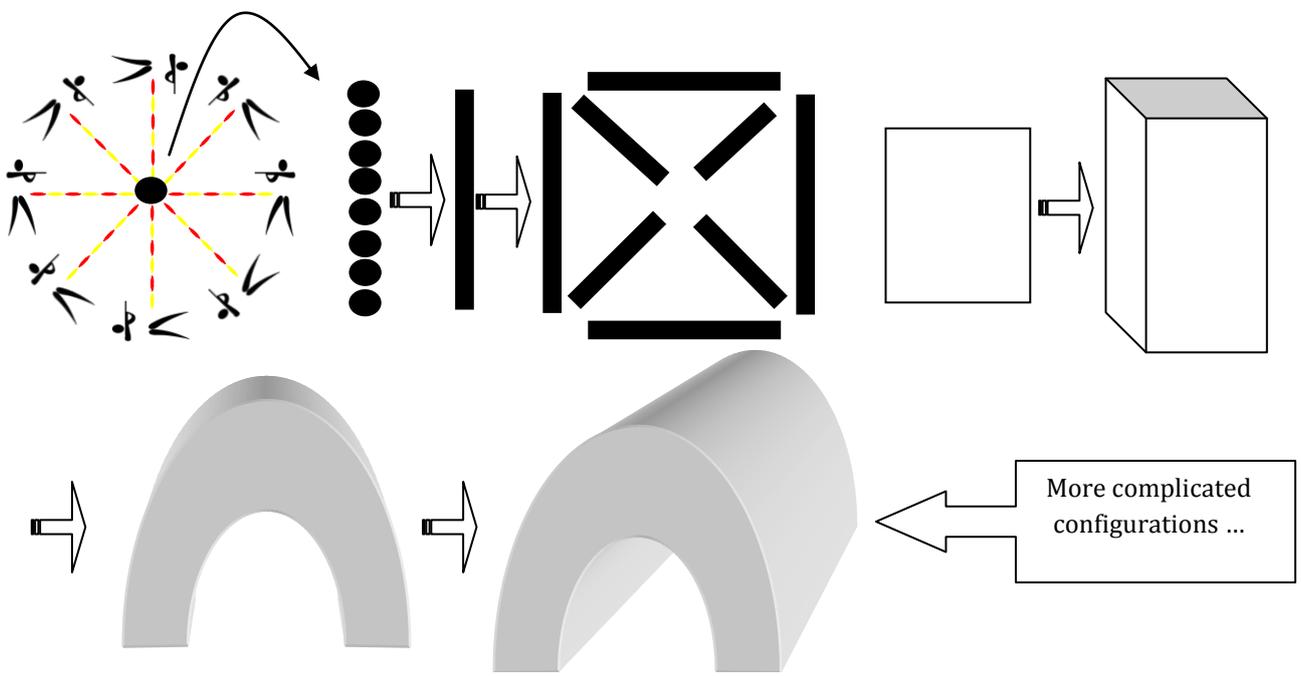

Figure 4: Following the Fig. 1, we set a collection of shooting the solitonized laser beams containing the squeezed regions of the vacuum. They represent a unique spot to interact and maximize the squeezing (in a steady state). The targeted spots can be developed in more complicated arrays. No explicit calculation has been given for this program, but the very idea is a generalization of the picked properties of the soliton theory.

One must note the demonstration of a continuous-wave Raman laser by Brasseur et al. [32] in experiments, has verified the predictions of [13].

**IV. The WDM Method**

In order to present a model for optical communication system of high-bit-rate data transmission in the nonreturn-to-zero (NRZ) format over transoceanic distance [29, see [25] too], considered as an alternative to the method of inverse scattering transformation [22] for the nonlinear Schrödinger (NLS) solitons [26], one observes the coded signal using the NRZ pulses. According to Fig. 1 of [29], for various types of a 16bits coded signal (0010110010111100):

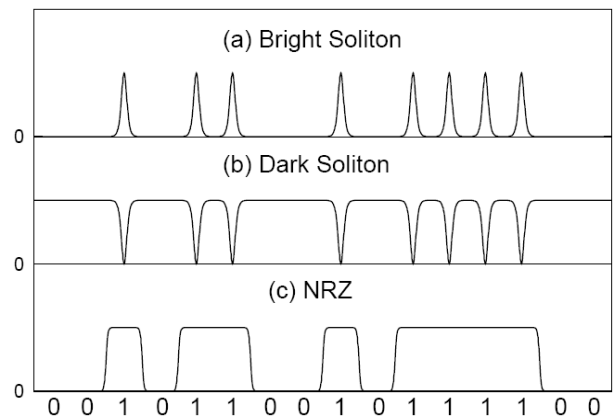

Figure 5

in the (011110) coding part, one has to operate the NRZ system at the defocusing regime, $\beta_2 < 0$, where $\beta_2$ represents the group velocity dispersion (GVD). We also might observe the antisymmetric dispersion managed (ASDM) solitons [18] in relation to the so-called twisted localized modes aka "dark-in-bright solitons" in Bose-Einstein condensates loaded in a periodic potential (optical lattice) [33] within the above figure.

The NRZ pulse is however non-stationary, and in order to reduce the pulse distortion, the applied limitation is designing the system in a small GVD and low power regime; therefore, if one makes an analogy to calculate the local intensity and phase of the laser field [34], it can be hoped to meet a compression of the pulse, i.e., expecting a shock. In a rather similar situation to NRZ, an optical shock appearing has also been noted in [35]. Therefore, while the wavelet analysis (see section **V**) is a mathematical tool in describing physical situations where the signal contains discontinuities and sharp spikes [36]; if the signal has sharp transitions, one can window the input data with the windowed Fourier transform (WFT) to localize the signal in time [37]. However, the physical demonstration of the used wavelets and WFT, depends on the experiments. The interested reader can refer to [29] to see the details on shock discussion – which we might need a chain of them – in the NLS-Whitham equation ($\frac{\partial \omega_1}{\partial Z} = \frac{\partial \omega_2}{\partial T}$, see section 3 of [29]).

Indeed, the initial phase modulation is set to be periodic with the period given by the width of one-bit pulse (bit-synchronous modulation), and the output is smoothed out by using a filter (averaged over high oscillations); Also, the two types of the solitons can be discriminated between by means of a simple wavelength filter [38]. On the other hand, one could establish the actual outcome of the gate (if one is to design a simple all-optical XOR logic system by PCSs[*]) by checking the content of the a field generated by overlapping between the colliding solitons called $U_4$ (which $u_n = \beta_1 U_n$ by Ref. [30] definition; see Eq. (9) too), in the output, by a suitable optical filter [30]; in fact, the guiding filters prevent the energies of solitons from decaying to zero in long (technically assumed, infinite) distances [39], but our model suffices to operate in rather short distances. Assuming the models as the nonlinear dynamical systems whose states are the energies of solitons, and by using techniques from the theory of dynamical systems, it can be shown it is impossible to maintain the energies of transmitted signals in different channels almost equal [39]. The differences of energies of signals can grow significantly large when the transmission length increases; thereby the problem of different energies of signals and decaying of some of them to zero in long distances can be overcome by guiding filters[ℓ] [39].

In a WDM system, signal in each channel has a different carrier frequency, and the model equation of the system can be obtained by setting the electric field $q$ of the NLS equation in the form $q = \sum_{j=1}^{N} q_j$; where $q_j$ denotes the electric field in the $j$-th channel with assumption of ignoring the mismatch of the frequency in the four wave mixing terms (FWM); and a DM technology is used to reduce the FWM and to keep a small GVD either. Furthermore, the Manakov equation (i.e., Eq. (6.2) of [29] with $\alpha = 1$) can be easily extended to an integrable N-coupled NLS equation which may correspond to a model equation of N-channel WDM system. Adopting limitations on the considered model of NRZ, one arrives at the following results: A system with $\alpha \neq 1$ may not be integrable ($\alpha$ is a normalization coefficient), in which as a practical example, one sets the equal intensity $\rho_1 = \rho_2 \equiv \rho_0$ and the opposite frequency shifts $u_1 = -u_2 \equiv u_0$ and $|u_0| > [(1 + \alpha)\rho_0]^{1/2}$ to obtain a minimum frequency separation[†] for 2-channel WDM system for a stable NRZ pulse propagation; in fact, those limitations are imposed to yield physically meaningful results.

In three papers of Malomed, et al. [38, 30, 18], where will be referred extensively in here, some soliton properties have been reviewed which can be interpreted usefully in relation to our purposes. The [38] studies on the solitons in a nonlinear optical fiber with a single polarization in a region of parameters where the fiber supports independently launching soliton streams in exactly two distinct modes, viz., the fundamental mode (FM) and the first-order helical mode (HM), as a controllable fashion to double the number of channels carried by the fiber, and despite the Kerr nonlinearity stating the linearly orthogonal solitons borne by the two modes interact via the cross-phase modulation (XPM), using a normalized system of coupled NLS equations for the interacting modes, computing nonstandard values of the XPM coupling constants in it, and results of "complete" and "incomplete" collisions[‡] between such solitons, the interaction between the two modes is additionally found to be strongly suppressed in comparison with that in the WDM system in the case when a dispersion-shifted or dispersion-compensated fiber is used. It is straightforward to note to the technical subtleties in our lasers arrangements, inspired by establishing an analogy between the results of [38] on the solitons in a different medium, namely the optical fiber. For instance, based on their distinct topological nature, FM and HM do not linearly mix[±], if the fiber remains circular, i.e., the fiber's bending will induce no linear mixing either, providing that the bending radius is much larger than the wavelength; Also, the standard waveguide parameter depends on the core radius, and the refractive index in the core and cladding, the full dispersion contains a material (bulk) contribution, as one observes the last two properties can be considered absent constraints in our model. We have to adopt appropriate magnitudes to admit soliton propagation in our laser beam, by taking into account direct observation of the polarization structure of subpicosecond solitons [10] in short fibers [40], very narrow solitary pulses in hollow nonlinear optical fibers [41], waveguide-geometry part of mode-dependent dispersion coefficients, lack of infinite integration intervals in some calculations, instability of the spatiotemporal spinning solitons ("light bullets") in bulk optical media (note: not in our picked up vacuum medium) with various nonlinearities[⊕] [42], and eventually implying sufficient mathematical simplifications.

It also is proposed that we would require long vacuum pumps plus cool-makers to surround the lasers. For typical solitons to be used in telecommunications [25], with the temporal width $T \sim 10$ ps, the Ref. [38] implies that a collision between the FM and HM solitons takes place at a propagation distance $z_{\text{coll}} \sim 500\ m$, which is much shorter than any soliton's length scale. Note that in the laboratory experiments with subpicosecond solitons[‡], the collision length may be $\lesssim 50$ m, implying that the experimental study of the collisions should be quite possible and hopes us that lacking intrinsic limitations of a fiber medium, we might be able to do the needed separations between the NE/PE pulses along such a distance by about 1000 times iterating of a solitonized beam shooting [10].

Furthermore, the collision-induced crosstalk, as the most fundamental problem in the soliton-based multichannel communication systems, between the FM and HM solitons – which may have different (pre-collision temporal) widths – is attenuated by the factor 0.62 for the HM soliton, as compared to the usual WDM system, while for the FM soliton the crosstalk strength is not different from the usual system, taking instead, the values around $V = 2.4$ (Fig. 1a of [38]), we will get small $\gamma_1 \approx 0.25$ , but large $\gamma_0 \approx 1.66$, which $V \equiv k\rho\sqrt{n_{\text{co}}^2 - n_{\text{cl}}^2}$ is the standard waveguide parameter and the effective XPM coefficient $\gamma_0$ and $\gamma_1$ of a normalized system of coupled NLS equations for the interacting modes, are given by the properly normalized overlapping integrals between FM and HM.

Having applied the helicity-generating phase masks, it is demonstrated that using two degenerate HMs with the opposite helicities S $= \pm1$, it is not really possible to triple the number of the channels; Also, doubling by means of the polarization-division multiplexing is incompatible with the WDM [43]. In addition, the mode's helicity is expected to be robust, as it is a *topological invariant*[±]. However, there is no group-velocity difference between the two envelopes of HMs with the helicities $\pm1$ (representing by $u_\pm$), hence the collision distance for the corresponding solitons may be very large, giving rise to a strong crosstalk between them. Moreover, the collision between two solitons with the helicities S $= \pm1$ may result in their annihilation or transformation into a pair of S $= 0$ solitons, while, due to the conservation of the topological invariant, the collision between the solitons with S $= 0$ and S $= 1$ is expected to be much closer to an elastic one. If one obtains the perturbation-induced evolution equations for the solitons' frequency shifts[∂], Eq. (13) in [38]:

$$\frac{d\omega_l}{dz} + \frac{4|\beta_l|\gamma_l}{T_{1-l}\sqrt{T_0^2+T_1^2}} \cdot \frac{d}{dt_l}\exp\left[-\frac{(t_l-t_0)^2}{2(T_0^2+T_1^2)}\right] = 0 \tag{7}$$

, it may furnish a closed dynamical system governing the evolution of the temporal positions $t_l$ of the two solitons. Now, something we attend, is the main effect of the complete collision[‡]: a "nonzero collision-induced *position* shift $\delta t_l$ of the soliton's center". Note that while the colliding solitons merge into a breather [44] in the usual system[†] [28], but those survive from the incomplete collision when they belong to the different modes, i.e., they pass through each other quickly enough, see Fig. 2 of [38]. Therefore, one concludes maybe the solitonization mechanism (section **III**) can be called to opting the related parameters which drive the lasing squeezed vacuum forward, like the soliton behaviors described above.

In [30] as a direct analog of WDMed models, parametric interactions in a quadratically nonlinear optical medium[φ] with the fundamental harmonic containing two components with (slightly) different carrier frequencies are studied. Indeed, a soliton may naturally be used as a bit of information (proper for the vital mechanism of guiding filters; see Fig. 5 too), and the interactions of solitons can potentially support logic operations (see section **V** too); in result the strongly inelastic collision between the orthogonal simple solitons, stable in a broad region, converts the simple solitons into polychromatic ones, and generating one or two additional polychromatic solitons (PCSs) generally having low power compared to simple solitons, but stable against small perturbations and playing the role of strong attractors in the system [28]; Equipped with a set of simplifying assumptions because of the serious difficulties with the experimental realization of the multi-resonance and/or multi-amplitude models [§], one finds the particular exact solutions of used system equations for stationary solitons as the form (see section II of [30]):

$$U_n = A_n\text{sech}^2(\lambda X) \quad, \quad n = 1,\dots,5 \tag{8}$$

Thus in the case of the full PCS, one has *two* nonlinear terms in the following equations:

$$i\frac{\partial U_1}{\partial Z} + \frac{\partial^2 U_1}{\partial X^2} - \alpha_1 U_1 + \chi_1 U_1^* U_3 + \chi_2 U_2^* U_4 = 0$$
$$i\frac{\partial U_2}{\partial Z} + \frac{\partial^2 U_2}{\partial X^2} - \alpha_2 U_2 + \chi_2 U_1^* U_4 + \chi_3 U_2^* U_5 = 0 \tag{9}$$

, rather than one term in the case of the simple solitons; Therefore, the necessary amplitude to compensate the spreading out of the beam due to the diffraction term is, roughly, twice as small in comparison with the simple solitons[§], or, eventually, the power is ∼ 4 times as small. The most significant inelastic effects occur when the solitons collide with the velocities ±0.4. In this case, the exit trajectory[*] is altered the most (see Fig. 6 of [30]), and, as per Fig. 7, the largest part of the net power is transferred into the newly generated harmonic components of the outer solitons[*]. Also the collision develops a phase difference across the fundamental-harmonic component fields and actually it is this (arbitrary) phase difference which attracts or repels the outer solitons, giving them non-zero velocities; there is another similar situation which through a direct simulation the two fairly robust-behaving pulses *repel* each other at an early stage of the interaction, and *attract* at a late stage [18], a desirable property to our approach. Hence, this collision may be used as a basis to design a simple all-optical XOR logic gate[ʊ]. An advantage of such a design of the XOR gate is that the post-collision state[*], i.e., any output beam will be a stationary soliton (even in the case of a strongly inelastic collision), which makes it convenient for further manipulations (cascadability).

On the other hand, since an increase in the incidence angle corresponds to an increase in the collision velocity, and angular spacing between the $U_1$ and $U_2$ amplitudes, as demonstrated in the Fig. 4f of [30], provides us more possibility of engineering, one concludes that the desired laser arrays in a 3-dimensional configuration would provide even more possibility of engineering consequently, from plate (2D) to space (3D). Moreover, by investigating the collisions between the ASDM solitons belonging to the different channels, one might estimate the crosstalk of Gordon-Haus timing jitter induced by collision between the solitons belonging to the different channels and find the timing-jitter suppression factor (JSF) for the ASDM soliton may be much larger than for its fundamental-soliton counterpart in the same system [18], also the energy of the stable antisymmetric soliton is four times as large as that for the fundamental DM soliton with the same width; in other words, the "heavier" soliton provides for better suppression of the timing jitter. Thereby, regarding the fact that the Hermit-Gaussian set of functions can be used to describe the propagation of pulses of a general shape in the strong-DM regime [18], the fundamental soliton of the Gaussian form being the first term in the set [46]; a system of ordinary differential equations may be derived by the variational approximation (VA), to describe the propagation of antisymmetric solitons in a multi-channel dense WDM optical fiber link – more useful for practical applications – with the kilometer and picosecond[‡] scales for length and time units respectively [18]; in dispersion-shifted fibers of optic telecommunications [25], having the pulse has zero chirp at the midpoint of each fiber segment, subjected to strong DM (i.e., periodic compensation of the GVD). Also, the position shift, along with a possible frequency shift[*] may be considered as the source of the timing jitter induced by the collisions, so for example the energy of the pulse $u(z, \tau) = A\tau \exp\left(-\frac{\tau^2}{W^2} + ib\tau^2 + i\emptyset\right)$, calculated according to the definition of conserving two channel energies $E_u \equiv \sqrt{2/\pi} \int_{-\infty}^{+\infty} |u|^2 d\tau$ , $E_v \equiv \sqrt{2/\pi} \int_{-\infty}^{+\infty} |v|^2 d\tau$ , is $E = A^2 W^3/4$, where $A$ and $W$ represent the amplitude and width of the pulse respectively. Thus, wider pulse and taller amplitude[§], more pulse energy. In addition, guiding filters[ℓ] are commonly used to suppress Gordon-Haus jitter and other noise effects in soliton transmission systems, they also lock the energies of solitons in different channels to fixed values that do not change with distance [39], causing error free soliton WDM transmission with positive steady-state energies in all channels, over transoceanic distances [47]. Alright, the seminal idea could be implying the same scenario of locking the energies of solitons over transoceanic distances, for "the solitons including the NED of the squeezed vacuum", as described at the solitonization mechanism. This idea surely requires more theoretical consideration, because its realizing would be far beyond my initial expectations to get exotic matter by the help of the solitons theory.

Observing the figures in [18] can be interpreted as a pattern to the engineering scenarios of [5, pp 10-11] for obtaining more NE; in which if one seeks for evidences looking like those waves in physics, these various samples can be some deserved candidates. Thereby, we arrive at a definite framework: To manufacture initial amounts of NE pulses out of the squeezed vacuum via solitons, then enhancing them to macroscopic amounts via solitons again! As an important required property for our scheme, a result points out "for a large propagation distance $z$, the position shift generated by the frequency shift*" grows as $\delta T_u^{(\omega)} = -\delta\omega\epsilon D_u z$ [18]; A position shift of $\delta T = 0.625$ has resulted from the complete collision‡; although this condition is not always met, as example for head-on complete collisions where the pulses are well separated, the frequency shift is negligible, whereas the position shift is significant, or in a case collision may completely destroy the ASDM solitons not predicted by the VA; additionally, while it is significant to see (Fig. 9 of [18]) that both the fundamental and antisymmetric solitons do not change their shapes after the collision; Anyway, analyzing the dynamics of solitons' collisions in [30, 18] is straightforward.

## V. Using wavelets

As a matter of fact, the information theory approach to displace the NEDs/PEDs is not a well-reviewed area of research. Apart from a hint in [6], this issue requires more review. The idea of compressing the data, moving them to a targeted location, then decompressing them, is physically attractive and mathematically possible; for example, that would be appropriate to compress the PE of the laser signal (as the tractable information) and after the NE packages did their job, to decompress the PEs as the heat, e.g., into the water.

To that end, the wavelet analysis provides immediate access to information that can be obscured by other time-frequency methods such as Fourier analysis [48], and since the chosen basis of adapted waveform carries substantial information about the signal [49], one might use some patterns, as an averaging smooth filter (see Eq. (10) and also to bring out the data's detail information [50], then one makes an inverse wavelet transformation to reconstruct the data set [36], (e.g., the Coiflet wavelet family (see Fig. 4 of [36], also [51, 52] for more knowledge) has a similar shape to a hyperbolic-secant soliton, intuitively resulting a better trade-off). This localization of the signal in time by windowing the input discretely sampled time-series data (consistent to squeezed light), along with wavelets' localization of frequency, makes many functions and operators using wavelets "sparse" when transformed into the wavelet domain. This sparse coding, in turn, results in a number of useful applications such as data compression, detecting features in images, and removing noise from time series.

Actually, wavelet transforms have an infinite set of possible basis functions which are useful at picking out frequencies and calculating power distributions that have to be soliton-shape in our framework. Hence, to span our data domain at different resolutions, the analyzing wavelet is used in a scaling equation:

$$W(x) = \sum_{k=-1}^{N-2}(-1)^k c_{k+1} \Phi(2x+k) \qquad (10)$$

which $W(x)$ is the scaling function for the analyzing wavelet $\Phi$, and $c_k$ are the wavelet coefficients. That would help one to consider the coefficients $\{c_0, \cdots, c_n\}$ as a filter [36]. According to Wickerhauser [49], five desirable properties for adapted wavelet bases can be listed as: 1. quick computation of inner products with the other basis functions; 2. fast superposition of the basis functions; 3. proper spatial localization, to identify the position of a signal that is contributing a large component; 4. suitable frequency localization, to identify signal oscillations; 5. independence, so that not too many basis elements match the same part of the signal. Eventually, it should be noted while one can apply the methods mentioned in Ref. [36] in our signal processing, including wavelet packet generators, envelope generators {e.g., see Eq. (5)}, etc; but the real manifestations shall be experienced in lab tests.

**Overview:** Three methods were given: 1. Developed amplification of the NEs along with dumping the PEs by means of the solitons. 2. Locking the NEs in guiding filters due to [39] analysis to transmit them to long distances. 3. Compressing the NEs or PEs and releasing them at somewhere else as its mathematical base can be worked via the wavelets.

There are eight alterable parameter in the present context: 1. Basic differential equations (NRZ, PCS, etc). 2. Shape differences among the solitons. 3. Amplitude differences. 4. Phase differences. 5. Frequency differences. 6. Velocity differences. 7. Mode differences (HM, FM, etc). 8. Angular spacing out.

Therefore, one concludes the optics and laser physics might provide new possibilities to generate the NED out of the vacuum, a more free approach than the confined Casimir effect [53], and it can be more engineered in result.

**Acknowledgement:** Thanks to Mali for the kind support; also to Wikipedia.org that helped me so much.

[10] However, intuitively speaking, one would imagine those proposals can be estimated effective as the initial chains of an enrichment process to give the macroscopic amounts of NED. In other words, the mirrors' setting of the Fig. 1 is expected to make more enriched light beam, i.e., "more negative" one, out of an incident beam, in order to deliver it to the other steps of the separation process. That can be reached practically in the regime of pulsewidth with scores of picosecond.

[16] Let alone depleting the vacuum, that is the case in more issues, such as collisions. For instance, when the energy is large or the group-velocity difference between the channels is small, which are detrimental features for the applications, the ASDM solitons may be completely distorted by the interaction [18], not quite useless for our purposes.

[21] One has to note since solitons are governed by non-linear evolution equations, a soliton can not be constructed by a linear superposition of waves.

[24] Having applied the soliton theory terminology, one needs to work on the self-phase modulation and dispersion term(s), broadening the pulse in the frequency domain or time domain, number of polarization(s), and so on.

[25] One advantage of our framework is, although solitons in fiber optic communication systems, where they may be harnessed to provide highly accurate signal transmission over extremely long distances [11], but we are not supposed to operate on such long scales that in result avoiding possible weakening in the mentioned medium. Regarding the Casimir effect implications, our chosen medium would be the vacuum, not the fiber.

[28] Three related examples are the NRZ pulse with an initial phase modulation that looks to tend to deform into an RZ pulse which may have a better property in signal processing in a network system of optical communication [29]. Along with polychromatic solitons which their low power causes them good agents to be engineered [30]; whenever necessary, after manipulating the base differential equations, they can be replaced with those solitons having more power. Also, the some DM pulses are not solitons in the strict mathematical sense, each one gets slightly distorted by the interaction [18].